\documentclass[10pt,preprint]{aastex}             
\usepackage{graphicx}
\usepackage{txfonts}
\usepackage{verbatim} 
\usepackage{natbib} 

\begin{document}
   \title{Polarization properties of real aluminum mirrors \\ I. Influence of the aluminum oxide layer}

   \author{G. van Harten, F. Snik, and C.U. Keller}
   \affil{Sterrekundig Instituut Utrecht, Princetonplein 5, 3584 CC, Utrecht, the Netherlands}
   \email{f.snik@astro.uu.nl}

  \begin{abstract}
   In polarimetry it is important to characterize the polarization properties of the instrument itself to disentangle real astrophysical signals from instrumental effects.
   This article deals with the accurate measurement and modeling of the polarization properties of real aluminum mirrors, as used in astronomical telescopes.
   Main goals are the characterization of the aluminum oxide layer thickness at different times after evaporation and its influence on the polarization properties of the mirror.
   The full polarization properties of an aluminum mirror are measured with Mueller matrix ellipsometry at different incidence angles and wavelengths.
   The best fit of theoretical Mueller matrices to all measurements simultaneously is obtained by taking into account a model of bulk aluminum with a thin aluminum oxide film on top of it.
   Full Mueller matrix measurements of a mirror are obtained with an absolute accuracy of $\sim$1\% after calibration.
   The determined layer thicknesses indicate logarithmic growth in the first few hours after evaporation, but it remains stable at a value of 4.12$\pm$0.08 nm on the long term.
   Although the aluminum oxide layer is established to be thin, it is necessary to consider it to accurately describe the mirror's polarization properties.
  \end{abstract}


%

\section{Introduction}

Polarimetry is a powerful astronomical technique for characterizing e.g. magnetic fields and scattering media.
The careful analysis of the polarization of light is, however, hampered by the instruments themselves as individual optical components induce and modify polarization.
To improve the polarimetric accuracy of current and future polarimeters, it is therefore crucial to know and understand the polarization properties of the telescope and the instrument.
All reflecting surfaces such as the aluminized telescope mirrors produce some instrumental polarization at oblique incidence angles.
The precise polarization properties of aluminum mirrors are notoriously difficult to model, because of variations of the refractive index, the presence of an aluminum oxide (Al$_2$O$_3$) film on the mirror, and the presence of dust on the surface \citep[e.g.][]{Harrington, Joos}.
This article deals with the empirical characterization and physical modeling of the non-ideal behavior of aluminum mirrors due to the aluminum oxide layer. The second paper in this series discusses the influence of dust on the surface (Snik et al. 2009; in preparation).

The importance of taking into account the aluminum oxide layer was theoretically shown by, e.g., \citet{Burge}.
Ellipsometry with a He-Ne laser was performed by \citet{Sankarasubramanian}, who showed the improvement of the Mueller matrix model after taking into account the aluminum oxide layer.
However, the deduced aluminum oxide thickness values ($\sim$ 50 nm) are an order of magnitude larger than what is expected from a layer formed by quantum mechanical tunneling of electrons through the layer at room temperature.
This theory, developed by \citet{Mott}, predicts logarithmic layer growth in the first few days, up to a thickness of about 5 nm.
This theoretical model successfully explained the thickness measurements performed by \citet{Jeurgens} based on electron microscopy and X-ray photoelectron spectroscopy, which yields an amorphous aluminum oxide layer with a uniform thickness of 0.5-4.0 nm for low temperatures ($\leq$ 573 K).
To fully characterize the aluminum oxide layer and its influence on an aluminum mirror's polarization properties as a function of wavelength and time after evaporation, we constructed a precision ellipsometer based on a broad-band light source in combination with interference filters.
Compared to the measurements of \citet{Sankarasubramanian} at a single wavelength and a single time after evaporation, we studied the growth of the oxide layer, by accurately measuring a mirror's Mueller matrices at different incidence angles, at several wavelengths, and at different points in time after evaporation.


\section{Measurements}

\begin{figure*}
  \centering
  \includegraphics{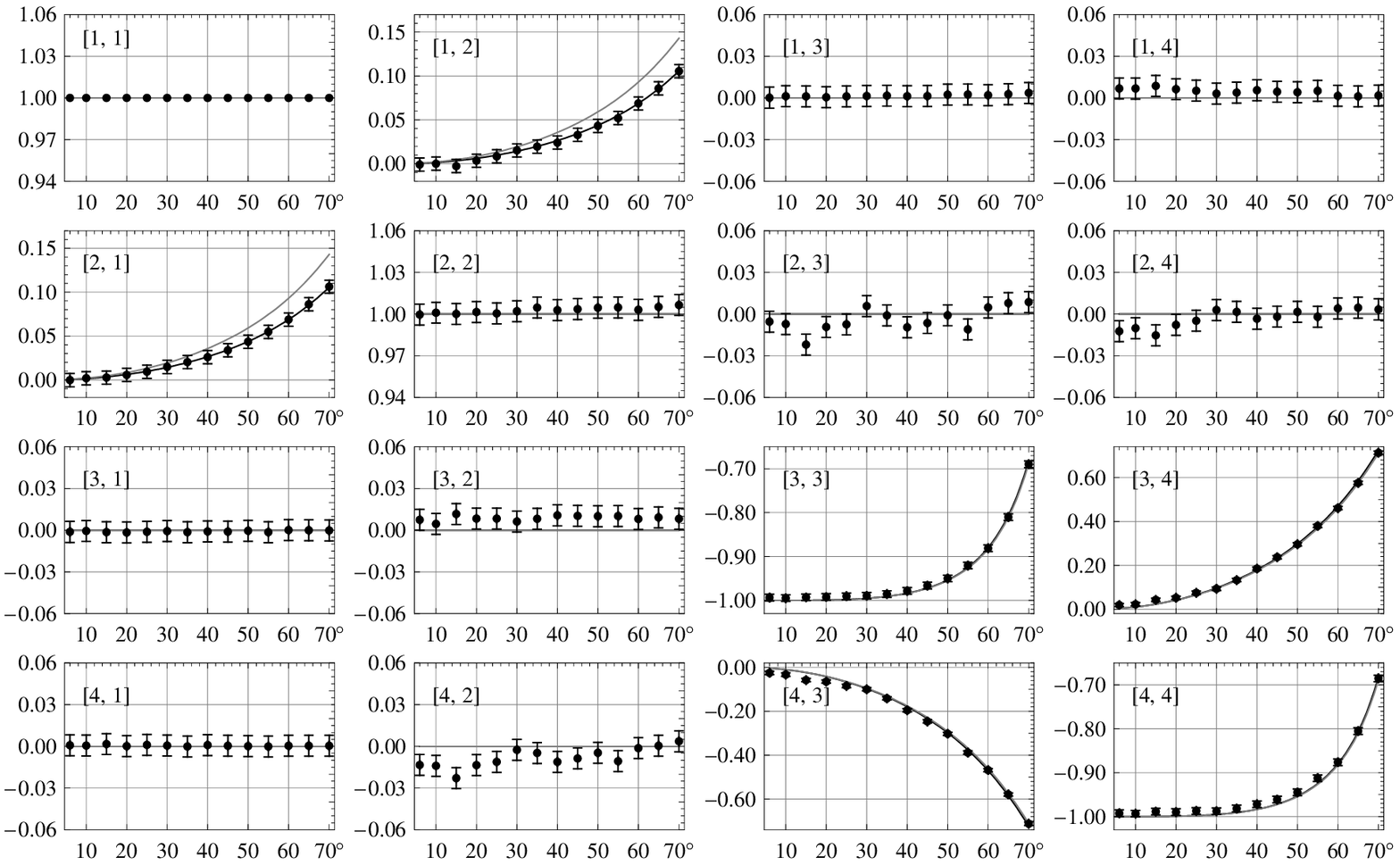}
  \newline
  \newline
  \includegraphics{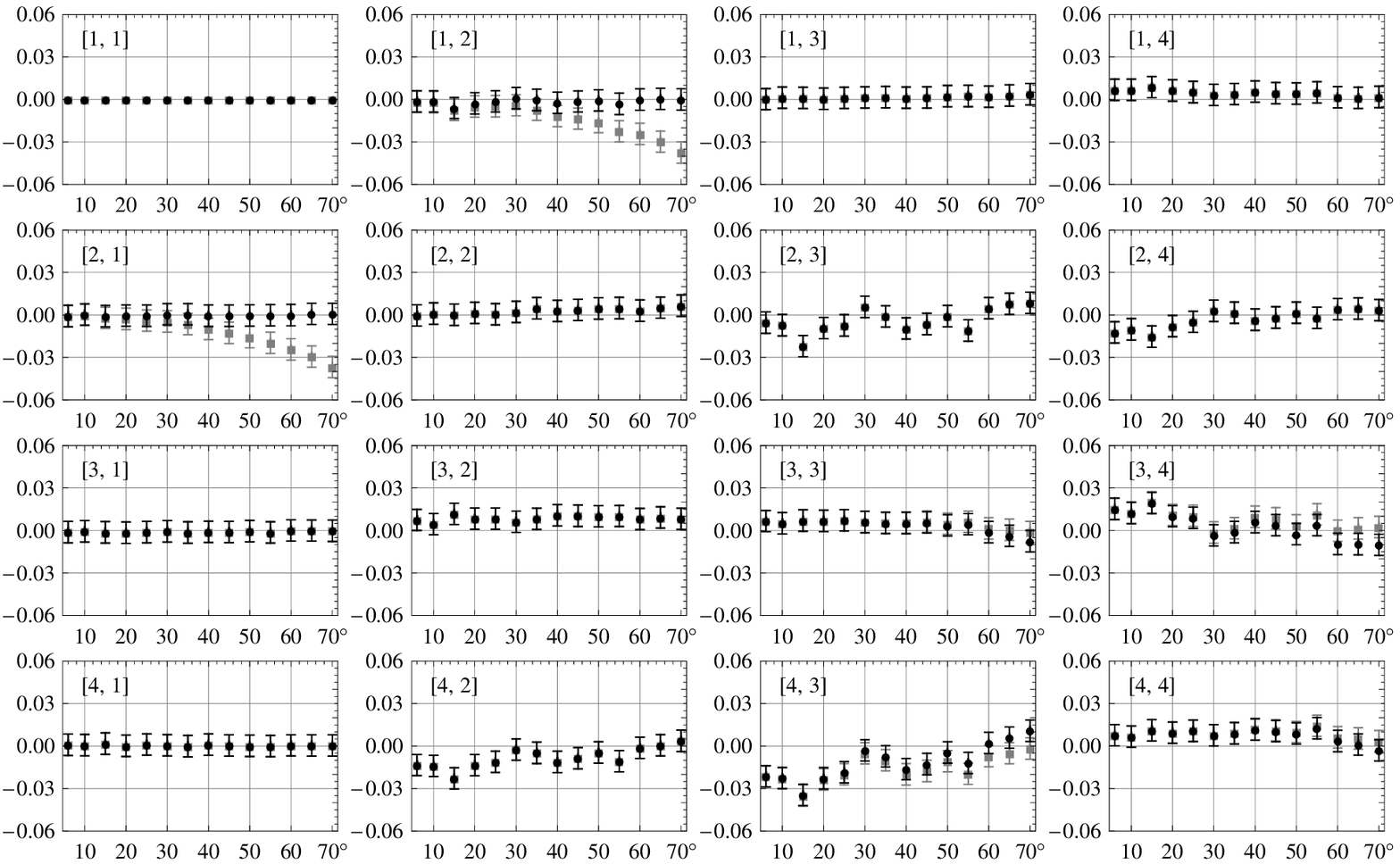}
  \caption{a. Mueller matrices of reflection off an aluminum mirror 768.8 hours after evaporation, at 600 nm.
  The solid curves are the least squares fits to a Mueller matrix model of reflection off aluminum with (black) and without (grey) a thin aluminum oxide layer on top of it. The best fit parameters for $k(\lambda=600)$ and $d$ (black), and for $k^{pp}(\lambda=600)$ (grey) are presented in Table~\ref{t2}.
  b. Residuals (data minus model) after curve-fitting, showing substantial, systematic residuals in the [1,2] and [2,1] elements if the aluminum oxide layer is not included in the model (grey squares).\label{f1f2}}
 \end{figure*}

A complete Mueller matrix ellipsometer, based on liquid crystal variable retarders (LCVRs), was developed for use in transmission, reflection and scattering measurements.
Any incidence angle of 6-70$^{\circ}$ can be achieved, and the angle of the measurement arm is independent of the incidence angle.
This instrument is able to characterize a sample's complete polarization properties in the form of a Mueller matrix by impinging light on it with any state of (fully polarized) elliptical polarization.
The Stokes vector of the emerging light is measured by a complete polarization state analyzer.
The sample's Mueller matrix is obtained after complete polarization measurements for at least four different input polarization states.
The Stokes coordinate system is chosen such that $\pm Q$ corresponds to vertical and horizontal linear polarization, respectively.
Positive Stokes $U$ is oriented at 45$^{\circ}$ clockwise from +$Q$ when looking along the beam.
$+V$ is oriented counterclockwise.
The polarization state generator (PSG) consists of a vertical polarizer ($0^{\circ}$), followed by a compensated liquid crystal variable retarder (LCVR) from Meadowlark Optics at $27.4^{\circ}$ (clockwise as seen looking downstream) and one at $72.4^{\circ}$.
Two LCVRs, the first at $-72.4^{\circ}$ and the second at $-27.4^{\circ}$, followed by a vertical polarizer, form the polarization state analyzer (PSA), which is indeed a mirror image of the PSG.
For every individual LCVR, the retardance curve as a function of voltage is unambiguously determined by measuring the transmission with the LCVR between parallel and crossed polarizers, with the fast axis at $45^\circ$ with respect to the polarizers, sweeping the voltage with steps of $50\textrm{ mV}$.
The modulation / demodulation scheme is given by the 16 retardance combinations of $1/8\lambda$ and $5/8\lambda$ at the LCVRs.
This set-up minimizes the error propagation to the determined Mueller matrix, as explained by \citet{DeMartino}.
To reduce the influence of systematic polarimetric errors due to imperfect components, the eigenvalue calibration method (ECM) by \citet{Compain} is implemented, which is the most accurate method currently known in the literature.
The ECM implies the determination of the system matrices, describing the PSG and the PSA, on the basis of measurements of a few reference samples with Mueller matrices with known parameter dependencies.
The reference samples are a Glan Thompson polarizer at $28^\circ$ and $73^\circ$, and a Soleil Babinet compensator with $\lambda/4$ retardance at $23^\circ$ and $68^\circ$, both used in transmission.
This choice is based on simulations of the calibration, showing that mutual angles of $45^\circ$ yield the most accurate results, and these exact angles maximize the accuracy by producing maximum output intensities, making the set-up least sensitive to noise and detector non-linearity.
In front of the PSG, a broadband halogen lamp, a collimating achromatic lens and a diaphragm produce a collimated beam, 1 cm in diameter, centered on the sample.
At the end of the PSA, the light is focused onto a photodiode detector.
A similar detector directly measures the lamp intensity at the PSG, such that lamp intensity fluctuations can be eliminated by normalizing all measurements of the PSA's detector to the reference detector.
Different interference filters are used to investigate the wavelength dependence of a sample's Mueller matrix.
The relatively low photon flux of the light source limited the choice of interference filters for this research to $\lambda$=500, 550, 600 and 650 nm, all with 10 nm band-pass.
Since we are primarily interested in the polarization properties of a sample and not in its photometric properties, all measured Mueller matrices are normalized to their [1,1] elements that only give a measure for its transmission or reflectivity.
With this set-up, a polarimetric sensitivity (i.e. the random noise level) of $2\cdot10^{-4}$ is achieved in the normalized Mueller matrix elements after a measurement taking 2 seconds per modulation / demodulation setting, at one incidence angle and wavelength.
The accuracy of the determined Mueller matrix per measurement is determined by the root mean square (RMS) of the elements that have a perfectly known theoretical value of 0 (and if possible also by the RMS of the deviations of the elements that are theoretically identical to the [1,1] element, i.e. normalized 1).
All measurements presented here exhibit an absolute accuracy of $\sim1\%$ of element [1,1], i.e. $\pm0.01$ per normalized Mueller matrix element, and the errors are mainly systematic as is shown in Fig.~\ref{f1f2}b.
This is likely due to the temperature sensitivity of the insulated LCVRs that are measured to be stable to within $0.25^\circ\textrm{C}$ during measurements and calibrations, which corresponds to a maximum retardance variation of 0.625 nm according to the supplier's specs.
As a consequence, the elements of the normalized system matrices vary up to 0.01.
The error propagation to the measured Mueller matrix fully depends on the Mueller matrix of the sample and the calibration components.
Note that fixed retardance deviations, due to limited LCVR calibration accuracy, are detected by the ECM.

220 $\pm$ 10 nm of aluminum were evaporated onto a $5\times5\textrm{ cm}$ square glass substrate under vacuum conditions ($1.33\cdot10^{-4} $ Pa).
The aluminum thickness is determined from step height measurements, performed at different locations on the mirror.
Hence it follows that the uncertainty in the thickness is a direct measure of the flatness of the surface.
The beam is assumed to be large enough to eliminate the effect of local roughness.
The size of the footprint changes as a function of incidence angle, but the mirror's macroscopic properties are also assumed to be uniform.
The characterization of the aluminum oxide layer was consecutively performed by means of reflection ellipsometry.
At five different intervals after evaporation, and at four wavelengths (see Table~\ref{t1} for observation and calibration details), the Mueller matrices of reflection off the aluminum mirror at 14 angles of incidence (see Fig.~\ref{f1f2}) were measured.
Both before and after the measurements at one wavelength, the ECM was performed to determine the system matrices.
Fig.~\ref{f1f2}a shows the normalized Mueller matrices at 768.8 hours after evaporation, at 600 nm, with the calibration performed after the measurements.
The top right and bottom left squares, together with element [2,2], determine the accuracy of the measurements represented as 1$\sigma$ error bars for all measured Mueller matrix elements, as described above.
Table~\ref{t1} shows all accuracies and the ellipsometer's stability over time, both long and short term, and over wavelengths.

\clearpage
\clearpage

\clearpage
 \begin{table}
  \caption[]{Absolute accuracy of the measured normalized Mueller matrices in percentages of element [1,1]. At every wavelength, the values with the calibration before and after the measurements are shown, respectively.\label{t1}}
  $$
     \begin{array}{rp{0.25 cm}|ccccccccccc|}
     \multicolumn{1}{c}{} & & \multicolumn{11}{c|}{Absolute$ $accuracy$ (\%)$} \\
     \hline
     \multicolumn{1}{c}{Time} & & \multicolumn{11}{c|}{Wavelength$ [nm $\pm$5]$} \\
     \multicolumn{1}{c}{$[hours $\pm$0.3]$} & & \multicolumn{2}{c}{500} & & \multicolumn{2}{c}{550} & & \multicolumn{2}{c}{600} & & \multicolumn{2}{c|}{650} \\
     \hline
      1.3 & & 1.16 & 0.88 & & 1.38 & 1.33 & & 1.15 & 1.40 & & 1.15 & 1.13 \\
     22.0 & & 1.39 & 0.95 & & 1.50 & 1.18 & & 1.36 & 1.18 & & 1.47 & 1.42 \\
     69.5 & & 0.87 & 0.82 & & 1.13 & 1.02 & & 1.07 & 0.92 & & 1.00 & 0.80 \\
    244.2 & & 1.03 & 0.96 & & 1.28 & 1.09 & & 1.11 & 1.01 & & 1.02 & 1.03 \\
    768.7 & & 0.80 & 0.77 & & 0.90 & 0.79 & & 0.88 & 0.75 & & 0.92 & 0.79 \\
    \hline
    \end{array}
  $$
 \end{table}
\clearpage


\section{Modeling}\label{sec:mod}

The measurements are interpreted with a theoretical Mueller matrix of bulk aluminum with a thin aluminum oxide layer on top of it.
This model consists of the usual reflection Mueller matrix and complex reflection coefficients \citep[see][]{Keller} from thin-film matrix theory \citep{Macleod, Born}.
It considers a bulk metal (subscript $b$) with an index of refraction $n_{b}(\lambda)=\tilde{n}(\lambda)-i k(\lambda)$ with $k(\lambda)\geq0$.
This minus sign follows from the fact that in the above thin-film calculations the electric field is proportional to $\exp[i\omega(t-\frac{n}{c}z)]$ instead of $\exp[i\omega(-t+\frac{n}{c}z)]$.
Hence it follows that a wrong sign makes $k$ a growing factor, while it should be a damping factor, because of conservation of energy.
On top of it a dielectric, amorphous thin film (subscript $f$) with thickness $d_f$ is located.
The surrounding medium is represented by subscript $m$, which in our case is air with $n_m = 1.00$.
The values of $\tilde{n}(\lambda)$ and the refractive indices of the aluminum oxide film ($n_{f}$) at the applied wavelengths are shown in Table~\ref{t2}.
The values of $k(\lambda)$ are widely varying throughout the literature \citep{Harrington} and are therefore assumed unknown and fitted along with the Al$_2$O$_3$ thickness $d_f$.

The reflection coefficients of p- and s-polarized light, as a function of the angle of incidence $\theta_0$, are given by

\begin{equation}
\label{eq:refl}
 r=\frac{\eta_m E_m-H_m}{\eta_m E_m+H_m} ,
\end{equation}

where

\begin{equation}
\label{eq:em}
 \begin{array}{l}
  E_m = \cos\delta_f+i\frac{\eta_b}{\eta_f}\sin\delta_f , \\
  H_m = \eta_b\cos\delta_f+i\eta_f\sin\delta_f ,
 \end{array}
\end{equation}

with

\begin{equation}
\label{eq:deltaf}
 \delta_f=\frac{2\pi}{\lambda}n_f(\lambda) d_f \cos\theta_f .
\end{equation}

The angles $\theta_f$ and $\theta_b$ can be calculated from the angle of incidence using Snell's law,

\begin{equation}
\label{eq:theta}
 \sin\theta_0=n_f(\lambda)\sin\theta_f , \\
 n_b(\lambda)\sin\theta_b=n_f(\lambda)\sin\theta_f \space.
\end{equation}

In the above formulae,

\begin{equation}
\label{eq:etap}
 \eta_{b,f,m}=n_{b,f,m}(\lambda)\cos\theta_{b,f,0} ,
\end{equation}

for p polarization, and

\begin{equation}
\label{eq:etas}
 \eta_{b,f,m}=\frac{n_{b,f,m}(\lambda)}{\cos\theta_{b,f,0}} .
\end{equation}

for s polarization, respectively.
The resultant normalized reflection Mueller matrix is given by

\begin{equation}
\label{eq:MR}
 \begin{array}{l}
  \mathbf{M_{R}}(n_{m}(\lambda),n_{b}(\lambda),n_{f}(\lambda),d_f,\theta_0,\lambda) = \\
  \qquad \left( \begin{array}{cccc}
  1 & \frac{-R_{p}+R_{s}}{R_{p}+R_{s}} & 0 & 0 \\
  \frac{-R_{p}+R_{s}}{R_{p}+R_{s}} & 1 & 0 & 0 \\
  0 & 0 & -\frac{2\sqrt{R_{p} R_{s}}\cos(\epsilon_{p}-\epsilon_{s})}{R_{p}+R_{s}} & -\frac{2\sqrt{R_{p} R_{s}}\sin(\epsilon_{p}-\epsilon_{s})}{R_{p}+R_{s}} \\
  0 & 0 & \frac{2\sqrt{R_{p} R_{s}}\sin(\epsilon_{p}-\epsilon_{s})}{R_{p}+R_{s}} & -\frac{2\sqrt{R_{p} R_{s}}\cos(\epsilon_{p}-\epsilon_{s})}{R_{p}+R_{s}}
  \end{array} \right) ,
 \end{array}
\end{equation}

where

\begin{equation}
\label{eq:refleps}
 \begin{array}{l}
  R_{p,s}=|r_{p,s}|^{2} \\
  \epsilon_{p,s}=$arg$(r_{p,s}) .
 \end{array}
\end{equation}

Weighted least squares fits are performed to determine the aluminum oxide thickness and the wavelength-dependent values of the imaginary part of the refractive index of aluminum for each time period.
Because $n_{f}$ is assumed to be sufficiently well known, and an extra degree of freedom causes unstable and unrealistic fit values, $n_{f}$ is not included in the fit (see Table~\ref{t2} for the literature values).
The sum of the quadratic differences between every determined Mueller matrix element and the corresponding element in the model is minimized with respect to $d_f$, $k_b(\lambda=500)$, $k_b(\lambda=550)$, $k_b(\lambda=600)$ and $k_b(\lambda=650)$.
The inverse of the accuracy per calibration determines the weight of the corresponding Mueller matrices in the fit.
As the errors of the determined values are difficult to establish analytically, Monte Carlo simulations are performed to determine the spread in the obtained $d$ and $k$.
Each Mueller matrix element (except element [1,1]) at each wavelength and incidence angle is randomly either increased or decreased with the accuracy value, after which the fit is performed again.
This is repeated 1000 times, and the standard deviations of the determined parameter values are used as 1$\sigma$ accuracies of $d$ and $k$, even though the measurement errors are not Gaussian.

It has been shown that the presence of a thin dielectric layer causes considerable changes in the polarization properties of a mirror \citep[e.g.][]{Burge, Sankarasubramanian}.
If the layer is neglected, the fit results in angle dependent pseudo-indices of refraction $n^p_{b}(\lambda)=\tilde{n}^p(\lambda)-i k^p(\lambda)$ that partly correct for the incomplete model.
To quantify these indices of this mirror, the fit and Monte Carlo simulations are also performed with the thickness set to 0, with $\tilde{n}^p(\lambda)$ and $k^p(\lambda)$ as free parameters.
Furthermore, the fit is performed with the imaginary part of the refractive index of aluminum as the only fit parameter (denoted by $k^{pp}(\lambda)$), in case one adopts the literature values of $\tilde{n}$, and the aluminum oxide layer is neglected.

Summarizing, the three types of performed weighted least squares fits are: $ $\\$ $
Fit 1. fit to $d$ and $k$, literature values of $\tilde{n}$ $ $\\$ $
Fit 2. fit to $\tilde{n}$ and $k$ (denoted by $\tilde{n}^p$ and $k^p$, respectively), no aluminum oxide layer in model $ $\\$ $
Fit 3. fit to $k$ (denoted by $k^{pp}$), literature values of $\tilde{n}$, no aluminum oxide layer in model. $ $\\$ $

\clearpage
  \begin{figure}
   \centering
   \includegraphics{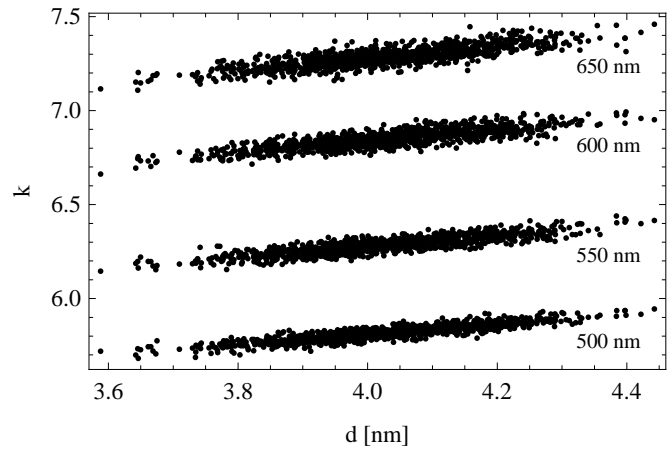}
   \caption{Result of Monte Carlo simulations of a 768.8 hours old mirror: 1000 fits of the data with random variations in $d$ and $k$ with an amplitude equal to the measurement accuracy (see Table~\ref{t1}). The standard deviations of $d$ and $k$ determine their accuracies.\label{f3}}
  \end{figure}
\clearpage


\section{Results and Discussion}

\begin{figure}
   \centering
   \includegraphics{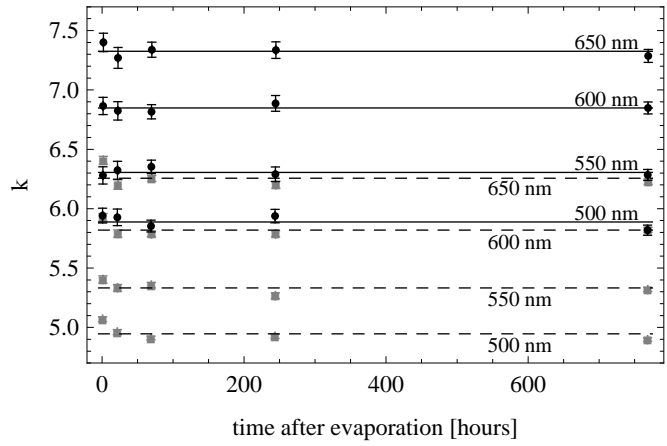}
   \caption{Fits for $k$ (black circles) and $k^{pp}$ (grey squares) at different wavelengths with and without aluminum oxide layer in the model, respectively. The lines are the weighted means per wavelength for $k$ and $k^{pp}$ (dashed) (see Table~\ref{t2}).\label{f4}}
 \end{figure}

In Fig.~\ref{f1f2}a one set of Mueller matrices is shown as well as the best fit for $d$ and $k$ for that day, with and without aluminum oxide layer in the model.
The residuals (Fig.~\ref{f1f2}b) show that with the literature value of the real part of the index of refraction of aluminum, with a pseudo-value of the imaginary part of the refractive index, the extended model is required to accurately describe elements [1,2] and [2,1].
The result of Monte Carlo simulations of the measurements of one day is shown in Fig.~\ref{f3}.
Despite the fact that the measurement errors are not Gaussian, the obtained distributions of $d$ and $k$ are inspected to have a distinct Gaussian shape.
It shows the error propagation to $d$ and $k$, and it shows that the parameters are correlated, but not to the level that a revision of $k$ can compensate for the aluminum oxide layer: its thickness $d$ is significantly larger than zero.
Fig.~\ref{f4} shows the determined values of $k(\lambda)$ and $k^{pp}(\lambda)$, as well as their weighted averages per wavelength.
It is clear that for Fit 3 (not taking the aluminum oxide layer into account) the determined values for $k$ significantly decrease with time, whereas the complete model (Fit 1) yields constant values for $k$.
Table~\ref{t2} shows the weighted averages of $k(\lambda)$ and the pseudo-index values, and $\tilde{n}^p(\lambda)$.
In agreement with \citet{Sankarasubramanian} and \citet{Joos}, the absolute values of all pseudo-indices are smaller than the actual values.

\begin{figure}
   \centering
   \includegraphics{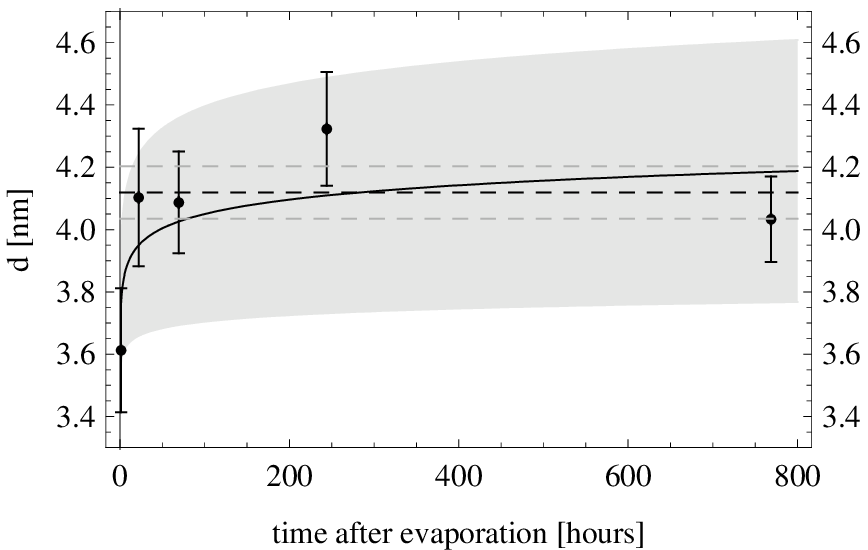}
   \newline
   \newline
   \includegraphics{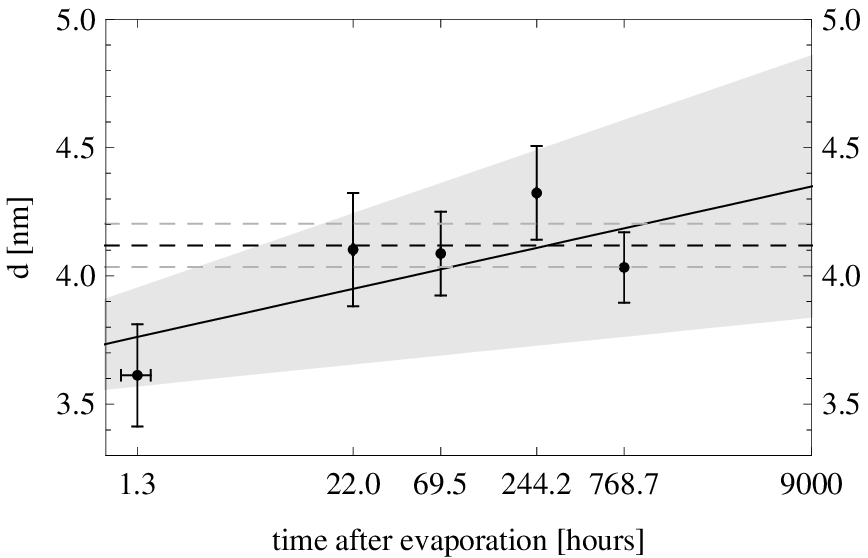}
   \caption{Results for the Al$_2$O$_3$ layer thickness as a function of time (linear and semi-logarithmic, respectively). The solid curve is the least squares fit to a logarithmic function, and the shaded region shows the error bars of the fit. The dashed lines show the weighted average and its error bars of the last four measurement points.\label{f5f6}}
  \end{figure}

The determined aluminum oxide layer thicknesses at different times after evaporation are shown in Fig.~\ref{f5f6} with linear and logarithmic axes for the time after evaporation.
The thicknesses are the best fit values for each measurement day, and the standard deviations therein are determined as the standard deviations of the simulated thicknesses per day, i.e. the standard deviation of $d$ at one wavelength in Fig.~\ref{f3} for each measurement day.
As the already present aluminum oxide prevents oxygen from reaching the aluminum, the layer growth is likely a logarithmic process \citep{Mott} starting out from small Al$_2$O$_3$ seeds.
In order to describe the Al$_2$O$_3$ layer thickness as a function of time, a least squares fit is made to the function $d(t)=a+b\ln(t)$.
The result, $a=3.74\pm0.18$ nm and $b=(6.6\pm3.6)\cdot10^{-2}$, is overplotted in Fig.~\ref{f5f6}.
The corresponding reduced chi squared is $\tilde{\chi}^{2}=1.26$.
As the growth has obviously reached a limiting layer thickness after the first day, the weighted average is also calculated without the first measurement point.
The result for the long term layer thickness is $d=4.12\pm0.08$ nm, with $\tilde{\chi}^{2}=0.56$.

The ten times larger thicknesses found by \citet{Sankarasubramanian} can possibly be explained by the applied values of the refractive indices of aluminum and aluminum oxide, including the sign of the imaginary part of the index of aluminum.
In fact, by applying the opposite sign of $k(\lambda)$, we obtained thicknesses of about 50 nm when fitting our data.

\clearpage
 \begin{table}
  \caption[]{Left: literature values of refractive index of amorphous aluminum oxide ($n_{f}$) and real part of refractive index of aluminum ($\tilde{n_{s}}$) at the applied wavelengths ($\lambda$).

  Right: real and imaginary parts of refractive index of aluminum, determined with three methods (see Section~\ref{sec:mod}): Fit 1: a fit of the measurements to $k$ and $d$ (literature values of $\tilde{n}$), Fit 2: to $\tilde{n}^{\mathrm{p}}$ and $k^{\mathrm{p}}$ (no aluminum oxide layer in model), and Fit 3: to $k^{\mathrm{pp}}$ (literature values of $\tilde{n}$, no aluminum oxide layer in model).\label{t2}}
  $$
     \begin{array}{c|cc|c|cc|c|}
     \multicolumn{1}{c|}{} & \multicolumn{2}{c|}{Literature} & \multicolumn{4}{c|}{Determined} \\
     \hline
     \lambda$ [nm]$ & n_{f}^{\mathrm{\dag}} & \tilde{n}^{\mathrm{\ddag}} & k & \tilde{n}^{\mathrm{p}} & k^{\mathrm{p}} & k^{\mathrm{pp}} \\
     \hline
     500 & 1.61 & 0.769 & 5.88\pm0.02 & 0.574\pm0.008 & 4.93\pm0.01 & 4.94\pm0.01 \\
     550 & 1.61 & 0.958 & 6.30\pm0.03 & 0.675\pm0.010 & 5.32\pm0.01 & 5.33\pm0.01 \\
     600 & 1.60 & 1.200 & 6.85\pm0.03 & 0.848\pm0.011 & 5.80\pm0.01 & 5.82\pm0.01 \\
     650 & 1.60 & 1.470 & 7.33\pm0.03 & 1.07\pm0.013 & 6.23\pm0.01 & 6.25\pm0.01 \\
     \hline
    \end{array}
  $$
   \begin{list}{}{}
    \item[$^{\mathrm{\dag}}$] $\pm$0.01, \citet{Eriksson}
    \item[$^{\mathrm{\ddag}}$] $\pm$0.01, \citet{Lide}, Linearly interpolated
    \end{list}
 \end{table}
\clearpage

\clearpage

\clearpage


\section{Conclusions}

We presented the results of Mueller matrix ellipsometry of reflection off an aluminum mirror.
Comparison with a thin film model shows the necessity of considering the amorphous aluminum oxide layer to accurately describe elements [1,2] and [2,1] of the Mueller matrix.
These elements are very important in astronomical polarimetry, because they describe the production of linear polarization (i.e. instrumental polarization).
Without the aluminum oxide in the model, this error is about 0.02 at a wavelength of $600\textrm{ nm}$ and an incidence angle of $45^\circ$ (see Fig.~\ref{f1f2}).
The cross-talk components $U\leftrightarrow V$ (elements [3,4] and [4,3]) can be represented by both the model with and without the oxide layer, because both $d$ (aluminum oxide layer thickness) and $k$ (imaginary part of the refractive index of aluminum) can affect the mirror's retardance.
This is also the reason why $d$ and $k$ are found to slightly correlate in Fig.~\ref{f3}: the presence of the layer can to some degree be accounted for by a decrease of $k$.
This is also observed in Fig.~\ref{f4}.
This has an interesting application in the sense that a mirror's retardance, and to some extent the induced polarization, can be modified by overcoating it with a dielectric layer \citep{Keller}.

The determined layer thicknesses at 5 different days after evaporation indicate logarithmic growth in the first few hours after evaporation, but it remains stable on the long-term at $\sim$ 4 nm, which is consistent with the theoretical model of \citet{Mott} and the experimental results of \citet{Jeurgens}.

With the obtained values of $k(\lambda)$ (see Table~\ref{t2}), $d=4.12\pm0.08\textrm{ nm}$ and the presented formulae (equations (\ref{eq:refl})-(\ref{eq:refleps})), a normalized Mueller matrix model of a realistic aluminum mirror can be constructed with an absolute accuracy of $\sim$1\%, i.e. $\pm0.01$ per normalized Mueller matrix element.
Our measurements were performed on a clean mirror, disregarding the influence of dust that is always present on telescope mirrors.
The polarization properties of a dusty aluminum mirror will be presented in a forthcoming paper (Snik et al. 2009; in preparation).


\acknowledgments
   The authors thank the reviewer, David Harrington, for his constructive and detailed comments on the manuscript.
   We also thank Jatin Rath for evaporating the mirror.


\begin{thebibliography}

\bibitem[{Born \& Wolf(1975)}]{Born}
Born, M., \& Wolf, E. 1975, Principles of Optics. Electromagnetic Theory of
  Propagation, Interference and Diffraction of Light, 5th edn. (Elsevier)

\bibitem[{Burge \& Bennett(1964)}]{Burge}
Burge, D.~K., \& Bennett, H.~E. 1964, Journal of the Optical Society of
  America, 54, 1428

\bibitem[{Compain {et~al.}(1999)Compain, Poirier, \& Dr\'evillon}]{Compain}
Compain, E., Poirier, S., \& Dr\'evillon, B. 1999, Applied Optics, 38, 3490

\bibitem[{{De Martino} {et~al.}(2003){De Martino}, Kim, Gracia-Caurel, Laude,
  \& Dr\'evillon}]{DeMartino}
{De Martino}, A., Kim, Y., Gracia-Caurel, E., Laude, B., \& Dr\'evillon, B.
  2003, Optics Letters, 28, 616

\bibitem[{Eriksson {et~al.}(1981)Eriksson, Hjortsberg, Niklasson, \&
  Granqvist}]{Eriksson}
Eriksson, T.~S., Hjortsberg, A., Niklasson, G.~A., \& Granqvist, C.~G. 1981,
  Applied Optics, 20, 2742

\bibitem[{Harrington \& Kuhn(2008)}]{Harrington}
Harrington, D.~M., \& Kuhn, J.~R. 2008, Publications of the Astronomical
  Society of the Pacific, 120, 89

\bibitem[{Jeurgens {et~al.}(2002)Jeurgens, Sloof, Tichelaar, \&
  Mittemeijer}]{Jeurgens}
Jeurgens, L. P.~H., Sloof, W.~G., Tichelaar, F.~D., \& Mittemeijer, E.~J. 2002,
  Thin Solid Films, 418, 89

\bibitem[{Joos {et~al.}(2008)Joos, Buenzli, Schmid, \& Thalmann}]{Joos}
Joos, F., Buenzli, E., Schmid, H.~M., \& Thalmann, C. 2008, Proceedings of
  SPIE, 7016, 70161I

\bibitem[{Keller(2002)}]{Keller}
Keller, C.~U. 2002, Astrophysical Spectropolarimetry, ed. J.~Trujillo-Bueno,
  F.~Moreno-Insertis, \& F.~Sanchez (Cambridge University Press), 303

\bibitem[{Lide(2008)}]{Lide}
Lide, D.~R., ed. 2008, CRC Handbook of Chemistry and Physics, 88th edn. (CRC
  Press / Taylor and Francis)

\bibitem[{Macleod(1969)}]{Macleod}
Macleod, H.~A. 1969, Thin-Film Optical Filters (Institute of Physics
  Publishing)

\bibitem[{Mott(1939)}]{Mott}
Mott, N.~F. 1939, Transactions of the Faraday Society, 35, 1175

\bibitem[{Sankarasubramanian {et~al.}(1999)Sankarasubramanian, Samson, \&
  Venkatakrishnan}]{Sankarasubramanian}
Sankarasubramanian, K., Samson, J. P.~A., \& Venkatakrishnan, P. 1999, Solar
  Polarization, ed. K.~N. Nagendra \& J.~O. Stenflo (Kluwer Academic
  Publishers), 313

\end{thebibliography}
\end{document}